\documentclass[doublecol]{epl2} 

\title{Reentrant transition of bosons in a quasiperiodic potential}

\author{A. Cetoli, E. Lundh}

\institute{Department of Physics, Ume{\aa} University, SE-90187 Ume\aa,
  Sweden}

\abstract{We investigate the behavior of a two dimensional array of
  Bose-Einstein condensate tubes described by means of a Bose-Hubbard
  Hamiltonian. Using a Wannier   function
  expansion for the wavefunction in each
  tube, we compute the Bose-Hubbard parameters related to two
  different longitudinal potentials, periodic and quasiperiodic. We
  predict that - upon increasing the external potential strength along
  the direction of the tubes - the condensate can experience a
  reentrant transition between a Mott insulating phase and the
  superfluid one.}

\begin{document}
\maketitle

\section{Introduction}
Since Greiner {\it et al.} first succeeded at storing a Bose-Einstein
condensate in a two-dimensional array of narrow tubes using an optical
lattice \cite{greiner2001}, experimental progress on low-dimensional
Bose systems has been tremendous. To name but a few highlights, the
Mott transition was observed in three \cite{greiner2002}, one
\cite{stoferle2004}, and two dimensions \cite{kohl2005,spielman2007},
and in 1D a Tonks-Girardeau gas has been realized
\cite{paredes2004,kinoshita2005}. Moreover, in 2D the
Kosterlitz-Thouless transition was observed
\cite{hadzibabic2006,schweikhard2007}.  The dimensional crossover
between three, two, and one dimensions for bosons in an optical
lattice was studied theoretically using Tomonaga-Luttinger liquid
theory in Refs.~\cite{ho,gangardt2006} and using Monte Carlo
simulations in Refs.~\cite{bergkvist2007, rehn2008}. These references
studied 2D or 3D optical lattices in which atoms can tunnel easily
along one Cartesian direction but not along the others. The system can
then be described as an array of tubes of bosons, which may or may not
be mutually phase coherent. As the tunneling between tubes is varied,
such a system will undergo a transition from a 3D superfluid (3D SF)
to a 2D Mott insulating phase (2D MI), which consists of a decoupled
array of 1D tubes. Similarly, if the tunneling probability along all
three Cartesian directions are made unequal, a 2D SF state can be
realized, in which there is superfluidity within separate 2D layers
but no coherence between them. These transitions are present only if
the tubes have finite length.

The theoretical approaches used in
Refs.~\cite{ho,gangardt2006,bergkvist2007,rehn2008} are able to describe phase
fluctuations within the tube-like filaments of bosons, but they do not
capture any nonlinear effects due to the possible variation in the
width of these tubes. Gross-Pitaevskii theory describes such
variations \cite{oosten}; however, it does so at the expense of not
being able to account for phase fluctuations.  Keeping these
limitations in mind, we offer in this Letter a description that is
complementary to those of Refs.~\cite{ho,gangardt2006,bergkvist2007},
using Gross-Pitaevskii theory in order to understand how nonlinear
effects may affect the phase transitions in this peculiar type of
optical lattice.

Such effects become important if the potential barriers along the
strongly coupled direction are so weak that each tube can be
considered as a quasi-1D Bose-Einstein condensate and the number of
bosons is large. It will be shown that these nonlinear effects can
give rise to a reentrant Mott transition in the array of 1D tubes.

The dilute Bose gas in an external potential $V_{ext}(\mathbf{r})$ is
described by the second quantized Hamiltonian
\begin{eqnarray}
\hat{H}[\hat\psi,\hat\psi^\dagger] = 
  \int d\mathbf{r}
   \Bigg[\frac{\hbar^2}{2 m}|\nabla \hat{\psi}(\mathbf{r})|^2
         &+& V_{\mathrm{ext}}(\mathbf{r})\,|\hat\psi(\mathbf{r})|^2 \\ \nonumber
	 &+& \frac{g}{2}\,|\hat\psi(\mathbf{r})|^4 
   \Bigg] \, .
\end{eqnarray}
The physical setting consists of a two-dimensional optical lattice in
the $x$ and $y$ plane with period $d$, creating a square
array of tubes which develop along the $z$ direction.  Moreover a
weaker optical potential parallel to the tubes is added, so that in
the present case the external potential is given by
\begin{eqnarray} 
V_\mathrm{ext}(x,y,z) = - V_x\,\cos\frac{2 \pi \,x}{d}
                  - V_y\,\cos\frac{2 \pi \,y}{d}
               + V_z(z)
	       \, .  ~~\label{extpot}
\end{eqnarray}

As in Ref.~\cite{jaksch} the many-body 
field operator
of the gas
$\hat{\psi}(\mathbf{r})$ is rewritten as a sum of local operators
\begin{equation} \label{wannier}
\hat{\psi}(\mathbf{r})= \sum_{j_x j_y k} 
                        \phi_{j_x j_y k}(\mathbf{r})\,\hat{b}_{j_x j_y k} \, ,
\end{equation}
where $\phi_{j_x j_y k}(\mathbf{r})=\phi_k(x+j_x \,d,y+j_y\,d,z)$
and each $\hat{b}_{j_x j_y k}$ acts on the $k^{th}$ state of the tube
in the position $(j_x,j_y)$. The $\phi_{k}$ are a complete set of
wavefunctions.  In the present setting we expect only the lowest
energy state to be occupied, so that it is possible to drop the $k$
index from~(\ref{wannier}).  Using the expression~(\ref{wannier}) the
second quantized grand potential
\begin{equation}
\hat{G}[\hat\psi,\hat\psi^\dagger]= 
  \hat{H}[\hat\psi,\hat\psi^\dagger] - \mu_{\mathrm{G}} \hat{N}[\hat\psi,\hat\psi^\dagger]
\end{equation}
can be rewritten
\begin{eqnarray}
\hat{G}
 = &-&\sum_{j} 
          \Big[ t_x   \, \hat{b}^\dagger_{j_x} \hat{b}_{j_x+1} 
                + t_y \, \hat{b}^\dagger_{j_y} \hat{b}_{j_y+1}
                + \mathrm{h.c.}\Big] \\ \nonumber
   &+& \frac{U}{2} \sum_j \hat{n}_j(\hat{n}_j-1)
   - \mu \sum_j \hat{n}_j 
   \, ,
\end{eqnarray}
where $j=(j_x,j_y)$, $\hat{n}_j=\hat{b}_j^\dagger \hat{b}_j$, and the
inter-tube tunneling matrix elements are
\begin{eqnarray} \label{tcoeff}
t_\alpha= &-& \int d\mathbf{r} 
     \, \Big[ 
          (\mathbf{\nabla} \phi_{j_\alpha})^* 
            \cdot \mathbf{\nabla}\phi_{j_\alpha+1} \\ \nonumber
          &+& V_\mathrm{ext}(x,y,z)\, \phi_{j_\alpha}^*\,\phi_{j_\alpha + 1}
        \Big]
\, ,
\end{eqnarray}
with $\alpha=\{x,y\}$. The in-tube interaction energy is
\begin{equation} \label{Ucoeff}
  U= g\, \int d\mathbf{r} \, |\phi(\mathbf{r})|^4 \, ,
\end{equation}
\begin{equation}
\mu= - \int d\mathbf{r} \, 
            \left[\frac{\hbar^2}{2 m} |\mathbf{\nabla} \phi(\mathbf{r})|^2 
                           + |\phi(\mathbf{r})|^2 \, V_\mathrm{ext}(\mathbf{r})
            \right] + \mu_\mathrm{G} \, ,
\end{equation}
where $\mu_\mathrm{G}$ is a constant that fixes the number $N_\mathrm{tot}$ of
particles in the whole system and $g=4\pi\hbar^2 a_s/m$. 

%Finding the wavefunction $\phi(\mathrm{r})$ is a nontrivial task: the
%problem can be reduced to retrieving an appropriate basis for the
%tubes, and then solving the many body ground state for this model.
%The basis should be chosen in a way that the many body ground state is
%as close as possible to the real one. The Hubbard parameters may then
%be computed using the knowledge of $\phi(\mathrm{r})$ in this basis.
%In principle the problem could be worked out by computing the ground
%state for a given basis using a quantum Montecarlo approach, and then
%minimizing the ground state energy among all possible basis. 
Finding the wavefunction $\phi(\mathrm{r})$ is a nontrivial task: 
We are ultimately concerned with approximating the full many-body ground 
state of the 3D system. In the absence of nonlinear on-site effects, 
the maximally localized Wannier functions can be used as a 
single-particle basis in which to expand the full many-body problem 
\cite{kohn,jaksch}. 
For the problem at hand, we are not only concerned with nonlinear 
effects from on-site repulsion, but also the elongated nature of the 
tubes. 
%EL
A rigorous recipe for how to choose the basis here cannot easily be 
formulated, but in principle the 
accuracy of the chosen basis could be assessed 
{\it a posteriori} by estimating how well the final solution approximates 
the 3D many-body ground state.  

%EL
Nevertheless, we take guidance from the 1D single-particle problem 
and apply the following 
\emph{ansatz} to describe the wavefunction in each tube,
\begin{equation} \label{ansatz_wann}
\phi(\mathbf{r})=  \sum_{i,j} a_{ij}(z)\,w_{i}(x)\,w_{j}(y)
\, ,
\end{equation}
where $\phi$ is normalized to unity
\begin{equation}
\int d\mathbf{r}\,\phi(\mathbf{r}) = 1
\,,
\end{equation}
and $w_i(x)$ and $w_j(y)$ are the 
Wannier functions 
for our lattice configuration along the $x$ and
$y$ direction respectively. 
Since we expect the ground state in each tube to be 
symmetric, the sum employs only the even Wannier functions. The energy
functional for the Bose-Einstein condensate in each tube is
\begin{eqnarray} \label{efun}
H[\phi,\phi^*]&=& \int d\mathbf{r} 
                \Big[ 
                     \frac{\hbar^2}{2 m}|\nabla\phi(\mathbf{r})|^2\\ \nonumber
              &~&~~~~~~  + V_{\mathrm{ext}}(\mathbf{r})\,|\phi(\mathbf{r})|^2 \\ \nonumber
              &~&~~~~~~  + \frac{g\,n}{2} \,|\phi(\mathbf{r})|^4 
                \Big]
%\nonumber\\
%             &=& E_{\mathrm{kin}} + E_{\mathrm{ext}} + E_{\mathrm{int}}
\, ,
\end{eqnarray}
with $n$ the mean number of particles in
each tube. Inserting
(\ref{ansatz_wann}) in Eq.\  (\ref{efun}), an expression of the total
energy as a function of the coefficients $a_{ij}(z)$ is obtained.  The
ground state has been found by minimizing this expression using the
conjugate gradient method; in the chosen set of parameters it is
enough to employ $0\le i,j \le 4$, since the occupation number of
higher Wannier states is negligible. Once the ground state is
obtained, the coefficients $t$ and $U$ can be computed by means of
(\ref{tcoeff}) and (\ref{Ucoeff}). 

\section{Periodic potential}

\begin{figure}[tbp]
\includegraphics[width=3.4in, clip]{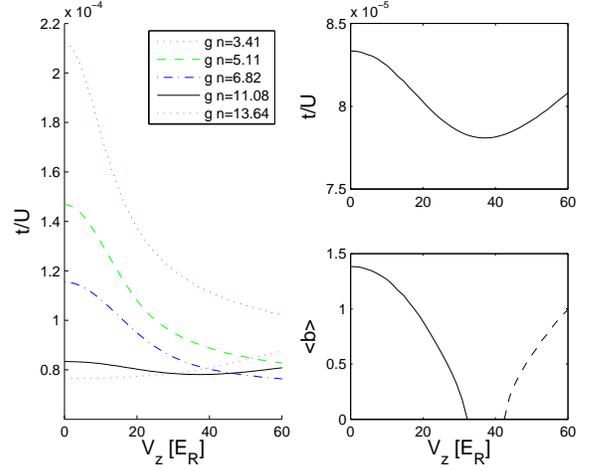}
\caption{Left panel: behavior of $t/U$ as a function of $V_z$ for some
  values of $g\,n$, using the the periodic potential (\ref{per_pot})
  with $V=77.35 E_\mathrm{R}$, as computed with the Wannier function
  approach.  Upper right panel: behavior of $t/U$ as a function of
  $V_z$ for $g\,n=11.08\,E_R$.  Lower right panel: tentative MF order
  parameter $\langle b \rangle$ against $V_z$ (see text) for the case
  $L=28$ ($11.9 \mathrm{\mu m}$), and $28$ particles for each
  tube.}\label{periodic}
\end{figure}

Let us suppose that the potential along the tubes is periodic. For
definiteness, let it have the same wavelength of the potentials in the
$x$ and $y$ directions
\begin{equation}\label{per_pot}
V_z(z)= V_z\,\cos \left( \frac{2\,\pi}{d}\,z \right)
\,.
\end{equation}
Focusing on the case in which $V=77.35 \,E_\mathrm{R}$ ($E_\mathrm{R}$
is the recoil energy $\hbar^2/m\,d$), in Fig.~\ref{periodic} (left
panel) we plot the behavior of $t/U$ while varying $V_z$ from $0$ to
$60\,E_\mathrm{R}$, for some values of the interaction strength. We
see that for $g\,n = 11.08 \, E_R$ the ratio $t/U$ decreases until
$V_z\approx 39 \,E_\mathrm{R}$, then it begins to rise.  It is known
\cite{fisher,sachdev} that the Mott insulating phase appears as a set
of ``lobes'' in the plane $\mu/U$-$t/U$, each one corresponding to a
precise number of particles $N_{\mathrm{tube}}$ for each tube; outside
of the 2D MI zone the system moves along lines of constant
$N_{\mathrm{tube}}$. The initial decrease and subsequent decrease of
$t/U$ therefore means that for an appropriate length of the tubes the
system can cross the 3D SF - 2D MI transition twice. In order to give
an estimate of the critical potential strength, the Gutzwiller
approximation is employed (see for example \cite{georges, sachdev}).
The condensate order parameter is given by the expectation value of
the destruction operator $\langle b \rangle$.  Figure~\ref{periodic}
(right lower panel) plots the behavior of this parameter for $L=28$
($11.9 \mathrm{\mu m}$) and $28$ particles in each tube.  The bounded
interval in which the superfluid phase vanishes is a clear consequence
of the nonlinear behavior of the system.

However, this example of a reentrant phenomenon does not survive a
more accurate analysis. In fact, the parameters are such that when the
system enters the 2D MI phase - thus isolating each tube from one
another - the gas of bosons is in an insulating phase \emph{within the
  single tube}. The assumption of coherence along the $z$ axis is not
valid anymore, and a reentrance to a 3D SF is questionable. Even
changing the parameters, we were not able to find a case where the
system remains superfluid along the tubes in the critical zone. This
situation, added to the experimental difficulty to obtain the precise
combination of parameters that realize the reentrant transition, makes
the described phenomenon very unlikely to be observed in the case of a
periodic potential along the tubes.

\section{Quasiperiodic potential}
\begin{figure}[tbp]
\includegraphics[width=3.4in, clip]{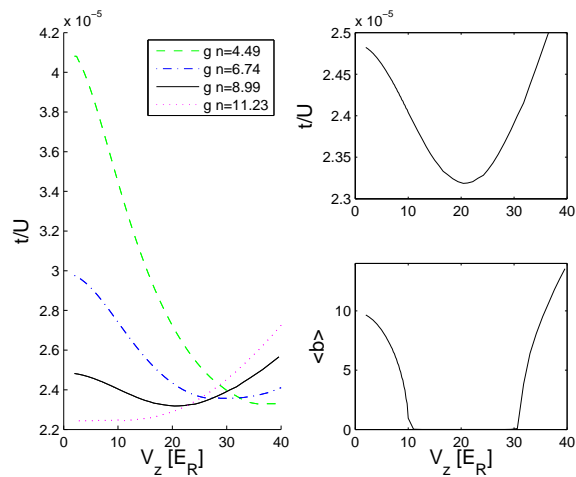}
\caption{Left panel: behavior of $t/U$ as a function of $V_z$ for the
  quasiperiodic case; the picture refers to some values of $g\,n$,
  while $V=100.0 E_\mathrm{R}$, as computed with the Wannier function
  approach.  Upper right panel: behavior of $t/U$ as a function of
  $V_z$ for $g\,n=8.99\,E_R$.  Lower right panel: MF order parameter
  $\langle b \rangle$ against $V_z$ for the case $L=89\,d$ ($37.825
  \mathrm{\mu m}$) and $2600$ particles per tube.}\label{qp}
\end{figure}
In order to avoid the Mott insulating phase within the tubes 
we employ instead a quasiperiodic
potential
\begin{equation}
V_z(z)= V_z\,\cos \left(\frac{2\,\pi}{d}\,z          \right) 
      + V_z\,\cos \left(\frac{2\,\pi}{d\,\lambda}\,z \right)
\,,
\end{equation}
where $\lambda = 89/55$ is an approximation to the golden ratio. We
have chosen to use the same lattice parameters as in the previous
analysis, with the length of the tubes chosen to be $L=89\,d$ ($37.825
\mathrm{\mu m}$). Proceeding in this way we see a behavior similar to
the one found for the periodic case: $t/U$ as a function of $V_z$ is a
non-monotonic function for $g\,n= 8.99\,E_R$, as shown in
Fig.~\ref{qp}. This situation gives rise to a reentrant transition
when the number of particles in each tube is between $2500$ and
$2700$.  Figure~\ref{qp} (lower right panel) shows the mean field
order parameter for $N_\mathrm{tube}=2600$. According to the
Gutzwiller approximation, in the shown parameter range
there are a maximum of 
$\langle b \rangle^2 \sim 100$ particles in the condensate.
\begin{figure}[tbp]
\includegraphics[width=3.4in, clip]{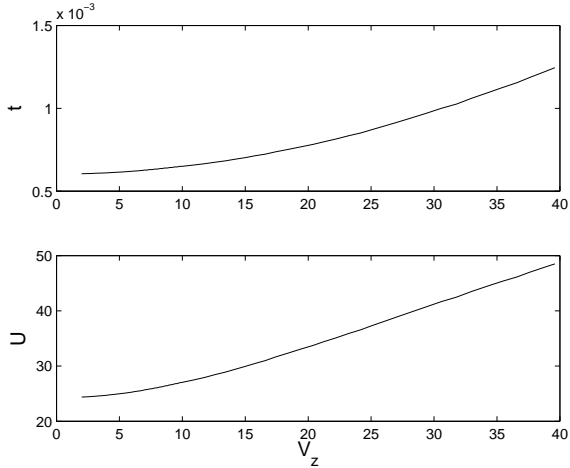}
\caption{Behavior of $t$ and $U$ as a function of $V_z$ for
  $g\,n=8.99 \, E_R$, in the quasiperiodic case. While they both grow upon
  increasing $V_z$, the different rate at which they increase is
  responsible for the nonlinear behavior of $t/U$
  (Fig.~\ref{qp})}\label{t_and_U}
\end{figure}
\begin{figure}[tbp]
\includegraphics[width=3.4in, clip]{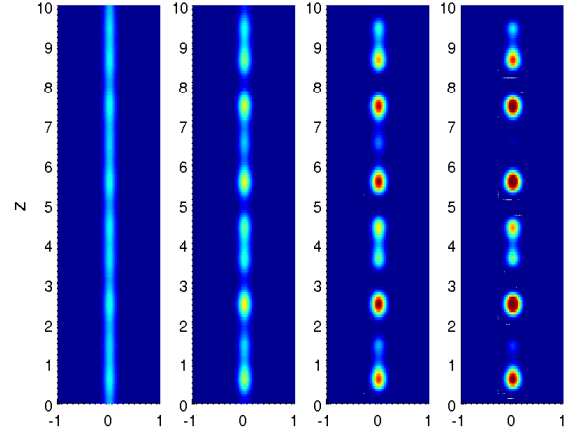}
\caption{Surface plot of the density profile $\int
  dy\,|\phi(x,y,z)|^2$ for - from left to right - $V_z=2, 10, 25$ and
  $40 \, E_R$, shown for $0 < z < 10\,d$. Increasing the longitudinal
  quasiperiodic lattice we see that the peaks in the boson gas profile
  become narrower, and at the same time the width of the wavefunction
  increases.}\label{phi_sq}
\end{figure}

In order to better understand the reason behind this phenomenon, the
plot in Fig.~\ref{t_and_U} shows the behavior of $t$ and $U$ against
$V_z$ separately for the case $g\,n= 8.99\,E_R$, and Fig.~\ref{phi_sq}
shows the density of the wavefunction $\int dy\,|\phi|^2$. Increasing
the longitudinal quasiperiodic lattice we see that the peaks in the
boson gas profile become narrower, thus raising the value of $U$
(\ref{Ucoeff}), while at the same time the wavefunction widens in the
radial direction, increasing the tunneling rate $t$ (\ref{tcoeff}).
At first $U$ rises faster than $t$ but for $t\approx 25
\,E_\mathrm{R}$ this relation is reversed.  The observed nonlinear
effect is therefore the result of a competition between $U$ and $t$,
i.e., between the increased density in a lattice minimum and the
broadening of the tubes upon raising the longitudinal lattice
strength.

In the presence of the quasiperiodic potential that we employ, the
Mott insulator phase within a single tube is inhibited, but a new
quantum phase comes along: the Bose glass. Recently a Bogoliubov
analysis has been employed to trace the boundaries of the Bose glass
phase\cite{fontanesi,cetoli_lundh}, and this method is expected to be
accurate within the limit of the Bogoliubov approximation, i.e., weak
interaction and high density ($g/n \ll \hbar^2/m$
\cite{leib_lininger}), as is the case for our current work. Using the
same method we were able to determine that the single tube never
leaves the superfluid phase during the reentrant phenomenon, thus
making it possible for the system to experience twice the 3D SF- 2D MI
transition while increasing the quasiperiodic lattice strength.

We thus conclude that our model is appropriate to predict the phase
transition, and the consequent reentrant behavior.

\section{Conclusions}
In this Letter we have studied the behavior of a 2D array of strongly
elongated tubes of bosons in an optical lattice, employing the
Bose-Hubbard Hamiltonian in order to describe the 3D SF - 2D MI
transition. Using a Wannier function ansatz and minimizing the
Hamiltonian of the system we were able to compute the ground state in
each tube, and link the parameter $t$ and $U$ of the Bose-Hubbard
Hamiltonian to the physical parameters of the atom gas.  Using this
approach we were able to see a nonlinear behavior of the parameters,
leading to the possibility of a reentrant transition between the 3D SF
and the 2D MI phase.

We applied our method to two different potentials along the tubes: a
periodic and a quasiperiodic modulation. While the non-monotonic
behavior of $t/U$ is seen - for a specific choice of the parameters -
in both cases, 
the periodic potential puts the tubes in a 1D MI phase, which would 
nullify the phenomenon we want to observe. 
In the quasiperiodic case, we predict 
that in an array of $^{87}$Rb tubes $37.825 \,
\mathrm{\mu m}$ long, with $d=425\,\mathrm{nm}$,
$V_x=V_y=100\,E_\mathrm{R}$, interaction strength $g\,n \sim 9\,E_R$,
and $N_\mathrm{tube}\sim 2600$, the system goes through the 3D SF - 2D
MI transition twice while increasing $V_z$. The system itself is found
to be in the insulating phase for $V_z$ between $\sim 10 \, E_R$ and
$\sim 30\,E_R$. We stress that this is not a consequence of a
localization phenomenon, 
but the quasiperiodic potential is only used to put the tubes in the 
right parameter regime. 

Future prospects include investigating how the algebraic decay of 
phase coherence -- which is not included in this study -- affects the 
detailed values of the critical parameters. 
%EL
Moreover, a real experiment will employ an inhomogeneous trapping 
potential. This introduces the complication that not all the
tubes would be equally long and the occupation number fluctuates. What
is more likely to be observed in this situation is a dip in the phase
coherence while varying $V_z$ in the range described above.
Finally, we expect the combined quasiperiodic-plus-periodic potential 
under consideration to have a rich phase diagram if a wider parameter 
regime is considered; this phase diagram will include MI in different 
dimensions, Bose glass, and combinations thereof. The investigation of 
these phases will be the subject of a separate study.

\end{document}